# Mimicking the Gas-Phase to Transport Odorants through the Nasal Mucus: Functional Insights into Odorant Binding Proteins

Massimiliano Paesani,[a],[b] Arthur G. Goetzee,[a] Sanne Abeln,[a],[c] and Halima Mouhib*[a]

Mammalian odorant binding proteins (OBPs) have long been suggested to transport hydrophobic odorant molecules through the aqueous environment of the nasal mucus. While the function of OBPs as odorant transporters is supported by their hydrophobic beta-barrel structure, no rationale has been provided on why and how these proteins facilitate the uptake of odorants from the gas phase. Here, a multi-scale computational approach validated through available high-resolution spectroscopy experiments reveals that the conformational space explored by carvone inside the binding cavity of porcine OBP (pOBP) is much closer to the gas than the aqueous phase, and that pOBP effectively manages to transport odorants by lowering the free energy barrier of odorant uptake. Understanding such perireceptor events is crucial to fully unravel the molecular processes underlying the olfactory sense, and move towards the development of protein-based biomimetic sensor units that can serve as artificial noses.

## Introduction

To trigger a signal in olfactory perception, odorant molecules and volatile compounds are bound to first travel by air through the nasal cavity, and subsequently cross the air-mucus interface to selectively bind their target odorant receptors (ORs) embedded in the membrane of the olfactory epithelium. During this process, odorant binding proteins (OBPs) are thought to facilitate the transport of hydrophobic odorants through the hydrophilic nasal mucus, and thus manifest the very first level of molecular mechanisms underlying the olfactory sense.[1–3] Although this provides us with a general understanding of odorant binding, the exact underlying mechanisms of action have not yet been reported at an atomistic scale. In particular, no attention has been paid to the intrinsic properties and the structural dynamics of the odorant molecules during the process. Little is known about relevant conformations that are sampled by the odorant molecule within the protein binding side. Here, we hypothesize that OBPs facilitate the transport of odorants through the air-mucus interface by mimicking the gas-phase environment through their characteristic hydrophobic binding pocket. To test our hypothesis and shed more light into the function of OBPs as odorant transporters, we focus on characterizing the conformational landscape explored by odorant molecules within the three different states: the gas phase, the aqueous phase, and the protein environment. The aqueous phase hereby represents the hydrophilic nasal mucus (simplified view). Using porcine OBP (pOBP) and its identified binding partner carvone (see Figure 1),[4] our approach targets studying the initial step in the olfactory process: the uptake of hydrophobic odorants from the gas-phase to the solvated phase. Hereby, we focus on how the conformational space sampled by the odorant is affected in these three environments. Understanding the conformations that are explored by the odorant molecule during this process provide understanding if and how OBPs facilitate their transport through the hydrophilic mucus.

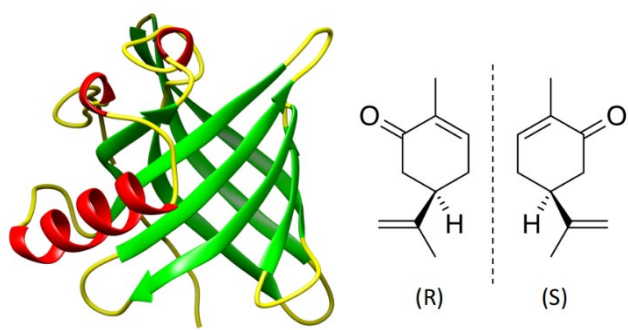

**Figure 1.** Left: Structure of porcine odorant binding protein (p-OBP, PDB-ID: 1A3Y[5]). The eight anti-parallel beta-sheets of pOBP are typical for members of the lipocaline family and form a strongly hydrophobic binding environment for small organic molecules. Secondary elements; beta-sheets, alpha helices, and loops are colored in green, red, and yellow, respectively.[6] Right: The two enantiomeric forms of carvone ((R)- and (S)-enantiomer).

To obtain atomistic detail into this process, a multi-scale computational approach is required to take all the different states into proper consideration. Here, stands the isolated odorant in the gas-phase as the initial step in the olfactory process, before subsequently moving into the solvated phase and the protein environment. A simplified overview of our muti-scale approach is provided in Figure 2. While the first state is best described at quantum chemical level, the latter two correspond to an increase in size of the molecular system of interest and require using classical molecular dynamics (MD) and enhanced sampling techniques.[7] To carry out the full conformational analysis of organic molecules such as carvone in the gas-phase, quantum chemical calculations may provide high-quality geometries that can be directly validated using high-resolution microwave spectroscopy experiments.[8] These rotational spectroscopy methods have previously been used successfully for conformational studies in fragrance research,[9] and are widely applied for the characterization of biomolecules[10–12] and complex systems[13–15] in the gas-phase. The unique experimental setup provides the advantage of studying the intrinsic properties of molecular targets in their isolated states in the gas-phase, by identifying different conformations in the gas.[16,17] This isolated state is the most simple state to study, before looking at solvation effects,[18,19] and the effects of the protein environment.[20] In this work, we validate the quantum chemical models of carvone directly with experimental data from high-resolution microwave spectroscopy experiments available in the literature.[21,22] The validated quantum chemical structures are then used to guide and set-up the MD simulations to quantify and characterize the conformational landscape of the odorant in the gas phase, in water, and within the pOBP binding pocket. We perform MD simulations of carvone *in vacuo*, in explicit water, and in complex with pOBP.[5] To guarantee their physical relevance, our protein simulations are guided using available experiments on the bound state of carvone and pOBP.[2,4] To further quantify the bound state of the protein-odorant complex, the free energy landscape of the carvone uptake through pOBP needs to be considered. Here, to quantify the stability and the free energy difference between the bound and unbound states, volume-based (VB) metadynamics provides a powerful enhanced sampling technique.[23] The setup and computational details of the production runs for all computational models at the different scales are provided in the electronic supporting information (SI).

[a] M. Paesani, A.G. Goetzee, Dr. S. Abeln, Dr. H. Mouhib*
Department of Computer Science, Bioinformatics, Vrije Universiteit Amsterdam, De Boelelaan 1105, 1081 HV Amsterdam, The Netherlands
E-mail: h.mouhib@vu.nl
[b] Van 't Hoff Institute for Molecular Sciences, Universiteit van Amsterdam, Science Park 904, 1090 GD Amsterdam, The Netherlands
[c] Department of Information and Computing Sciences, Department of Biology, Utrecht University, Heidelberglaan 8, 3584 CS Utrecht, The Netherlands

## Results and Discussion

### *Carvone Adopts Two Sets of Distinct Conformations in the Gas-phase*

Prior to looking at the structure and dynamics of carvone inside the protein complex, the conformational landscape of the molecule needs to be characterized without any influence from the environment. This step describes carvone in its isolated state in the gas phase using high level quantum chemical methods. To this aim, several starting conformations of carvone were generated for the optimizations at quantum chemical level through rotation around the $sp^3$ hybridized bonds, including chair and boat conformations of the carbon cycle. The input structures were subsequently optimized using the hybrid exchange-correlation functional of Becke-Lee-Parr (B3LYP)[24,25] from density functional theory, and the post-Hartree Fock method Moller-Plesset perturbation theory of second order (MP2)[26] in combination with the 6-311++G(g,p) Pople triple-zeta basis set. The quantum chemical computations yielded six stable conformers with relative energy difference below 10 kJ/mol for each carvone enantiomer, which were verified using harmonic frequency calculations to exclude transition states (see SI for full details). From a structural point of view, the six different conformers can be clustered into two distinct subsets based on the orientation of the isopropenyl group: one subset with an equatorial orientation (hereafter referred to as subset I) and another subset with an axial orientation of the isopropenyl moiety (hereafter referred to as subset II). The structural superposition of the *(R)*-carvone conformers into subset I and subset II is shown in Figure 3. Note that the six conformers of carvone have been labeled as conformer 1, 2, 3, 4, 5, and 6, according to their increasing relative energies from the the global minimum (conformer 1) as obtained at the MP2/6-311++G(d,p) level of theory. Subset I and II each comprise 3 of the six conformers optimized at quantum chemical level (subset I: conformers 1, 3, and 5; and subset II: conformers 2, 4, and 6, see Figure S2). Note that within each subset, the three conformations differ by the rotation of the $sp^3$ hybridized rotatable bond between the carbon-cycle and the isopropenyl group (see Figure 3). Using high-resolution microwave spectroscopy experiments, two stable conformers were identified in the gas-phase spectra by Moreno *et al.*[21] under molecular beam conditions. Both experimentally observed conformers exhibit an equatorial orientation of the isopropenyl group and correspond to structures found in subset I (see Figure 3). The equatorial orientation of side groups and substituent is often energetically favored over axial orientations, unless strong intramolecular interactions favor the latter. The reported experimental rotational constants are in good agreement with the geometries optimized at the MP2 level of theory (below 1% deviation) and the most abundant energy conformer observed in the experiment corresponds to the global minimum identified at quantum chemical level using both, the MP2 and B3LYP method. It should also be noted that enantiomers possess the same rotational constants and can therefore not be distinguished in the reported rotational spectroscopy experiment.[21] However, for the sake of completeness and also to validate the computational biology models, we included both, *(R)*- and *(S)*-carvone throughout all computational methods of our study. The optimized structures of the lowest energy conformer at the MP2 level of theory serve as starting point for the molecular dynamics simulations of both enantiomers of carvone in explicit water, to mimic the hydrophilic nasal mucus, *in vacuo*, and in the protein binding cavity. The full overview of the quantum chemical results including the Cartesian coordinates, relative energies and charges of all optimized conformers for both carvone enantiomers is provided in Section I of the SI.

### *pOBP Remains Unchanged upon Binding*

Before looking into the the bound form of the carvone and pOBP protein-ligand complex (holo form), the structure and dynamics of pOBP in its free and unbound state (apo form) need to be considered. Like most mammalian OBPs (with the exception of bovine OBP)[28], pOBP forms a stable monomer in solution. Subsequently, the holo form including one carvone enantiomer at a time can be tackled. Overall, the beta-barrel structure of pOBP is extremely stable, and experiments have shown that even truncated mutants missing the first eight residues of at the N-terminal maintain their stability in solution.[29,30] To analyse the stability and monitor the flexibility of the monomeric protein, we performed molecular dynamics simulations using the AMBERff99SB-IDLN force field implemented in GROMACS (version 2020.3). Altogether, 10 explicit water simulations of 50 ns were run for each set-up: pOBP in its apo-form and two holo-forms, i.e., in complex with either the *(R)*- or the *(S)*-carvone enantiomer. A detailed description of the simulation set-up is given in section 2 of the SI. From the MD production runs, the root mean square deviation (RMSD) allows us to monitor the overall changes of the pOBP structure and showed that there show no difference between the holo-form and the apo-forms of pOBP. In addition to the RMSD, we also define tree cross-barrel distances between opposing residues around the center of mass of the protein as specific order parameters to observe any changes of the binding site (see SI Figure S5). Again, there is no difference between the apo- and holo-forms of pOBP (see Figure S4 and S6). This means that the overall structure of the binding cavity is not impacted by the presence of the ligand throughout the simulations. These observations are in agreement with observation from experiments, where it was shown through several studies that mammalian OBPs are not likely to undergo any conformational changes upon ligand binding.[29,30] Putting this is perspective of the putative transporter function of OBPs, we note that the binding cavity of pOBP has a volume of approximately 500Å$^3$ [31], while a carvone enantiomer will take around 200 Å$^3$ of space (estimating 18 Å$^3$ for each non-hydrogen atom). There is thus sufficient room for carvone to adapt different poses and conformations inside the binding pocket without impacting the overall protein structure. This is further supported by Xray data of several pOBP-odorant complexes, and supports the putative transporter function of OBPs in the nasal mucus,[5,31] providing the opportunity to study the conformational space explored by carvone within the protein binding pocket.

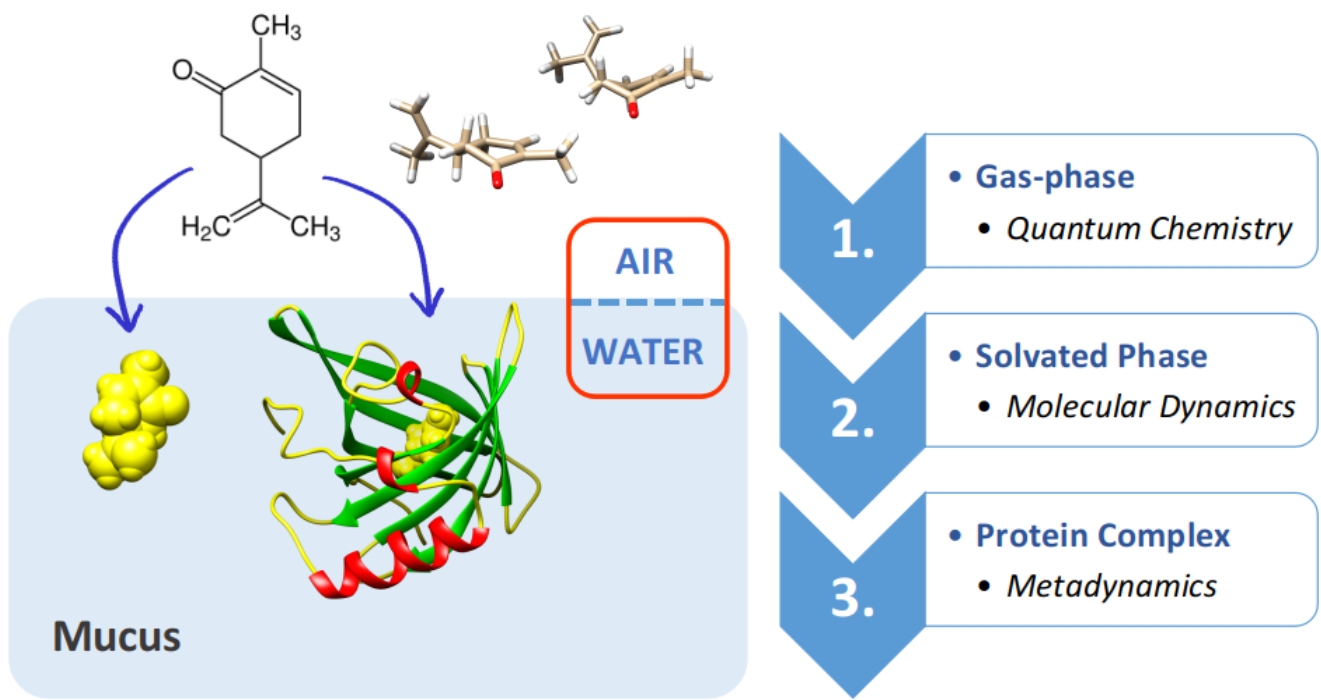

**Figure 2.** Overview of the multi-scale approach implemented in this work to study the transfer of carvone from the gas- to the aqueous phase, and in its bound state with pOBP. This approach allows to compare the conformational space that is explored by carvone in the different environments using adapted computational techniques. The computational set up for each scale is provided in the SI.

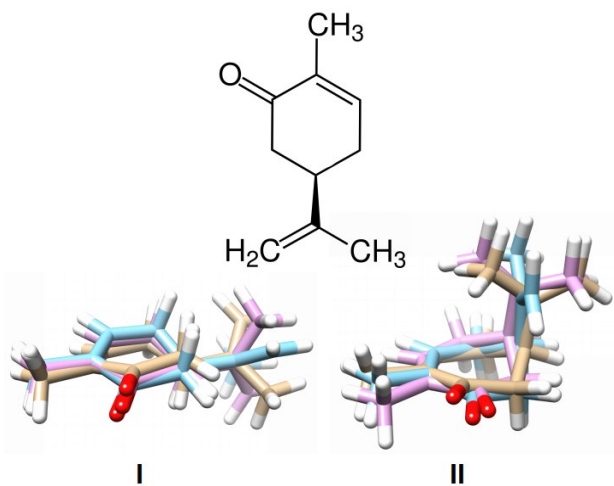

**Figure 3**. Structural Superposition of the optimized conformers of *(R)*-carvone at the MP2/6-311++G(d,p) level of theory using the UCSF Chimera program package. Note that the six energy minima can be classified into two distinct subsets of conformations with an (I) equatorial, and (II) axial orientation of the isopropenyl moiety. All optimizations were carried out using the GAUSSIAN16 program. package[27] The conformers clustered within subset I and subset II correspond to the conformers 1, 3, 5 and to the conformers 2, 4, 6, respectively (see Figure S2 for details). The full overview of the quantum chemical results is provided in section 1 of the SI.

## *Conformational Sampling of Carvone Shows that pOBP Provides an Ideal Hydrophobic Environment to mimic the Gas Phase*

In addition to the simulations of the pOBP-carvone complex, both enantiomers of carvone were simulated using molecular dynamics in explicit water and under *in vacuo* conditions to compare their structure and dynamics in all three states: gas (*in vacuo*), mucus (explicit water), and protein environment (pOBP-carvone complex). Since the predicted energy differences between the six different conformations of carvone are estimated to lie below 10 kJ/mol at quantum chemical level, both conformational subsets shown in Figure 3 are likely to be observed in the gas-phase simulations. Within a different environment such as water or inside the protein, however, the conformational space is likely to be confined to subspace due to the influence of the environment. For instance, the hydrophobic interface of carvone is likely to be minimized as a results of the hydrophobic effects. It is known that the curvature of small solutes affects the hydrophobic effect of the molecules, and one conformational subset of carvone could be favored over the other depending on the environments.[32,33] To monitor the conformational landscape of carvone during the simulations, we define a molecular distance $d$ between two carbon atoms (one in the isopropenyl group and one in the hydrocarbon ring, see Figure 4) as an intrinsic order parameter. This allows us to distinguish the conformational state of carvone, i.e., whether it is sampling the states in subset I or sampling the states in subset II. Figure 4 depicts the order parameter $d$ during the interconversion of conformer 1 (representing the structures of subset I) to conformer 6 (representing the structures of subset II) through the connecting transition state optimized at the MP2/6-311++G(d,p) level of theory. This order parameter $d$ is ideal to explicitly distinguish between the two different subsets of carvone. It is certainly more insightful than the RMSD for such a small molecular system, since the RMSD may be ambiguous to distinguish between the different conformations. Next to the distance $d$, the radius of gyration $R_G$ may also serve as a useful measure since the curvature of carvone is clearly different between the two subsets (see Figure 3 Section 2 of the ESI).

Throughout 10 MD simulations of 50 ns using randomized starting velocities for each production run, both enantiomers of carvone are sampled in the gas and inside the protein environment, allowing us to estimate the free energy barrier between the two states. In explicit water however, the barrier of inter-conversion between the structures in subset I and subset II of carvone is too high and the timescale of the simulations does not allow proper sampling between the different states. Therefore, enhanced sampling techniques are needed to accelerate the simulations and determine the barrier height in explicit water. Here, we apply umbrella sampling (US) which is a powerful technique to estimate the relative free energy of different states along a defined reaction coordinate. Using conformer 1 (representing the structures in subset I) and conformer 6 (representing the structures in subset II) as the start and end point of the US simulation, respectively, and the distance $d$ as reaction coordinate to link the two states, we were able to estimate the barrier to interconversion between the two carvone subsets I and II. The corresponding free energy plots in all three states, *in vaccuo*, in explicit water, and inside the protein environment, are shown in Figure 4. In the gas phase, both conformational ensembles are sampled, although a preference can be observed towards the conformations of subset I, with an equatorial orientation of the isopropenyl group. This is in agreement with the quantum chemical calculations, which show that these conformations are energetically more favorable than subset II, which contains conformers of axial orientation. Although the energetically favored state remains unchanged in the explicit water simulations, the barrier height is significantly higher than in the gas phase and inside the protein binding cavity. In the gas phase and inside the binding cavity of pOBP the barrier height is significantly lower, which allows carvone to fully explore its conformational landscape. This indicates that the protein environment is able to mimic the gas phase, and we hypothesise that this *gas-like environment* is what allows hydrophobic odorants to cross the interface between the gas and the hydrophilic mucus.

## *Characterizing the Lowest Energy State: Insight into the Free Energy Landscape of the Carvone-pOBP Complex*

From these findings we hypothesize that the bound protein-odorant complex is energetically favored and that this can be directly linked to the transporter function of pOBP. To simulate the unbinding process of carvone and pOBP, and estimate the change in free energy between the bound and the unbound state, the application of enhanced sampling techniques is required.[34] Enhanced sampling techniques such as metadynamics, steered molecular dynamics, or umbrella sampling allow to characterize binding affinities within the protein complex. These methods usually rely on the intrinsic molecular properties of the problem at hand, such as the unbinding pathway, to properly define the reaction coordinate around which the free energy landscape should be sampled. However, although putative binding pathways and affinities have been reported for several OBPs including pOBP (see Table 1),[1,3,35] these findings cannot be directly applied to the carvone-pOBP complex, as they, e.g., focus primarily on studying the binding affinity from selected mutants of pOBPs for chiral discrimination.[2,4] Since no clear unbinding pathway has been reported for carvone and pOBP in the literature, we cannot define reliable collective variables (CVs) for our system. In such as case, volume based metadynamics (VB-MD), a recent extended version of funnel-based well-tempered metadynamics)[36], is the method of choice to estimate the free energy of binding of our system, as it does not require prior knowledge of the unbinding pathway.[6,23,37] Unlike funnel metadynamics,[38] which samples the free energy along a predefined reaction coordinate, VB-MD samples all possibilities of the ligand to escape from the protein binding site within a defined sphere until it reaches the unbound state.[23] This method is adapted to identify numerous unbinding pathways of a protein-ligand complex and estimate relative free-energy difference that are comparable to experimental values. Figure 4 shows the resulting free energy landscapes for the unbinding of *(R)*- and *(S)*-carvone projected on two intuitive CVs: the distance of carvone from the center of mass of the protein ($\rho$ in nm), and the coordination number (CN), which captures the amount of non-covalent contacts between the respective carvone enantiomer and the residues of the pOBP binding site. These two CVs allow to get an accessible de-scription of the molecular structure underlying the protein-odorant complex. A detailed description of VB-MD, the definition of the CVs, and details of the simulation set up including the free energy landscapes for different radii of the sphere (< 28 Å), are provided in the SI. The relative free energy of the bound and unbound states can be obtained by subtracting the energy values of the global minima (the bound state) by the energy of the unbound states (cf. Table 2). Hereby, the unbound state can be defined somewhat arbitrarily as described by Capelli *et al.*[23]. We thus define the unbound state in the range of $\rho > 2$ Å and a low CN (CN < 10). The bound state on the other hand can be found in the region close to the center of mass $250 < CN < 350$ and $\rho > 0.5$ Å. The structure of two states and their relative locations on the free energy landscape are depicted in Figure 4. In agreement with experimental observations, we observe very similar contours for the free energy landscape of *(R)*- and *(S)*-carvone, indicating indeed that the wild-type pOBP does not distinguish between the two enantiomeric forms.[2,4] This is in alignment with our hypothesis that odorant binding proteins serve as transporters helping odorants to cross the air-mucus interface. As such, they should be non-selective to be able to transport a maximum number of odorants towards the olfactory receptors in the biological medium.

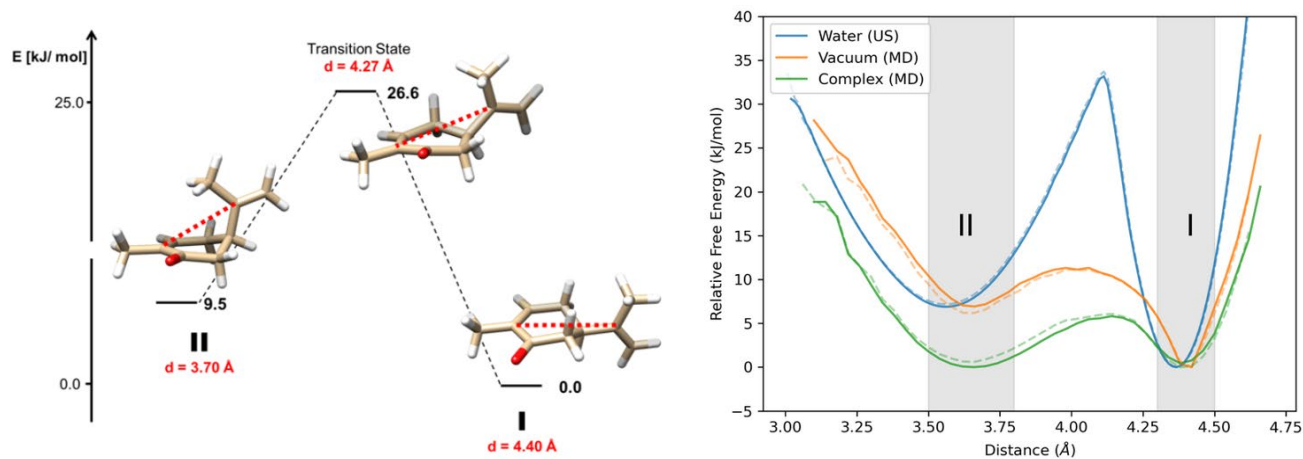

**Figure 4.** Left: Reaction coordinate for the interconvertion of *(R)*-carvone from conformer 1 (global minimum and representative conformer for subset I, see Figure 3) through the joint transition state to conformer 6 (representative conformer for subset II, see Figure 3). The distance between the two carbon atoms used as order parameter *d* to distinguish the different conformational states of carvone in the gas phase is highlighted in red. The structures and relative energies in kJ/mol were obtained after optimization at the MP2/6-311++g(d,p) level of theory; Right: Free energy surfaces of *(R)*-carvone in 3 different phases: water (Water), gas (Vacuum), and protein environment (Complex), the free energy surfaces of *(S)*-carvone are depicted as dotted lines. US and MD: umbrella sampling simulations and molecular dynamics simulations, respectively. See SI for further details.

## Conclusion and Outlook

### *Characterizing the Driving Force of Odorant Transport*

Using pOBP and carvone as a molecular targets, we set up a multi-scale bottom-up approach to elucidate how OBPs are able to facilitate the transport of odorants through the hydrophilic nasal mucus. Our hypothesis is that due to their strongly hydrophobic beta-barrel binding cavity, OBPs facilitate the transport of hydrophobic odorants though the nasal mucus by lowering the energy barrier at the air-mucus interfaces. This is achieved by mimicking the gas-phase environment inside the binding pocket, thus allowing the odorant to explore its full conformational flexibility inside the protein. Our multi-scale approach is tailored to simulate the monitor the conformational landscape of the odorant carvone throughout the different states it undergoes during the olfactory process: in its isolated state in the gas-phase, in the solvated state, and within the protein environment. Each state requires different computational methods and while the gas phase is best described at quantum chemical level, the solvated phase and the protein environment necessitate computational biology methods such classical molecular dynamics (MD) and enhanced sampling techniques. The conformational sampling at quantum chemical level revealed that carvone exhibits six stable energy minima that can be grouped into two distinct structural clusters (named structural subset I and II, cf. Figure 3) that are distinguishable through the orientation of the isopropenyl moiety of the molecule. This allowed us to identify the intramolecular distance *d* as an intrinsic order parameter to distinguish the different structures within the two subset throughout our simulations. To monitor the protein dynamics of the free pOBP, cross barrel distances and the RMSD were used. Our results showed that the structure of pOBP is remains unchanged when carvone is bound inside the beta-barrel and no strong interactions could be identified between carvone and protein residues within the binding cavity. As a result, the odorant retains the freedom to sample all the conformations within its conformational space. This complies with the low specificity of OBPs who need to be able to bind a high number of different hydrophobic molecules to fulfill their transporter function. The MD production runs further showed that the conformational space sampled by carvone in the gas-phase during the *in vacuo* simulations strongly resembles the conformations it samples inside the binding cavity of the protein. In both cases, the barrier barrier height for the interconversion is below 10 kJ/mol. In contrast, the barrier height for the interconversion between the different conformers is more than three times higher in explicit water (see Figure 4). This strongly indicates that the hydrophobic environment provided by the protein allows the odorant to maintain its conformational freedom upon binding and directly suggests that the protein environment is more air like. VB-MD additionally show that the free energy of the bound state is lower and favors the uptake of the odorant into the hydrophobic cavity. Given that carvone, as most other odorants, is hydrophobic and almost insoluble in water (cf. Table 1), our findings emphasize that OBPs primarily facilitate the uptake of odorants through the mucus layer. Still, this does exclude the existence of parallel transport mechanisms where odorants may dissolve directly into the hydrophilic mucus and diffuse towards the ORs. However, this alternative route is likely to be kinetically hindered supposing that the purpose of having OBPs in the mucus is to speed up the transport and ultimately the detection of the olfactory signal at receptor level. This view is supported by experiments showing that the olfactory response can also be initiated in the complete absence of OBPs.[39] However, the receptor response is observed at reduced efficiency and sensitivity, which complies with the idea that effectively less odorants will be able cross the air-mucus interface in the absence of OBPs. Finally, being members of the lipocalin super family, it is possible that OBPs take an active part in additional tasks such as a scavenger role during the termination of the olfactory signal,[40] protecting ORs from over saturation, or to particpate in the nasal immune response by binding to pathogens or their products.[41]

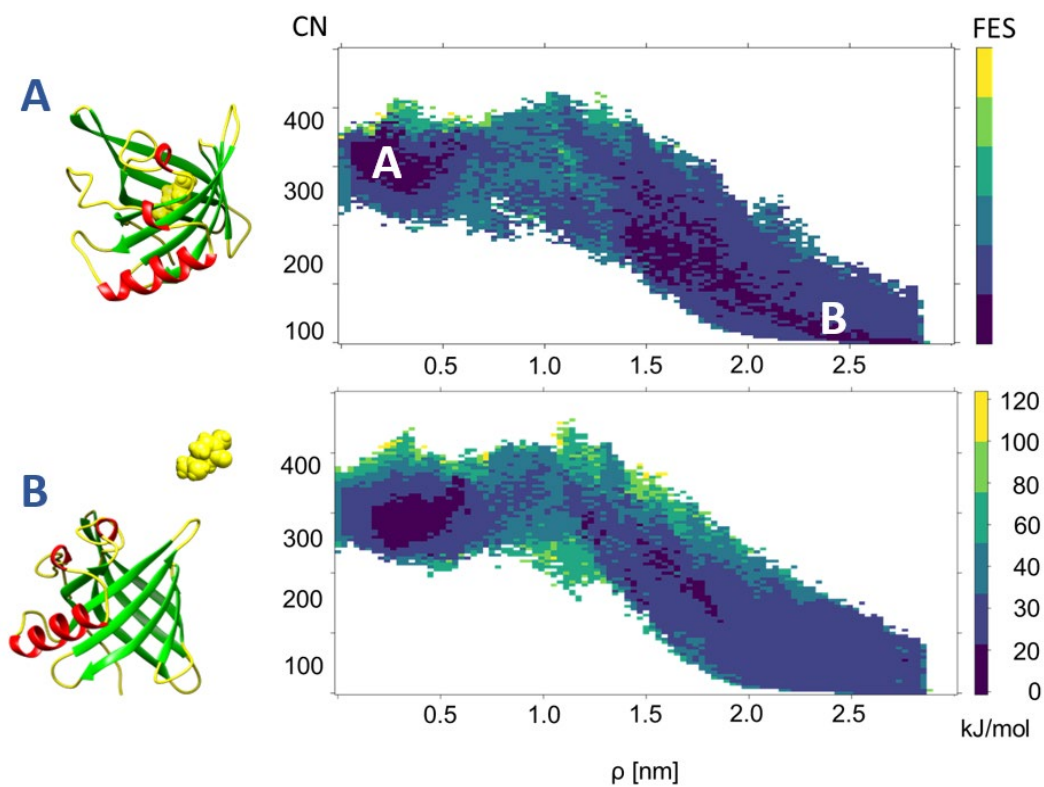

**Figure 5.** Free energy surface of the pOBP carvone complex generated using volume-based metadynamics for a $\rho$ of 28 Å (Upper trace: *(R)*-carvone. Lower trace: *(S)*-carvone)). CN: coordination number (measure for the number of contacts between the ligand and the protein binding pocket), A and B: bound and unbound states of the pOBP-carvone complex, respectively. See SI for all technical details and the simulation setup.

**Table 1.** Odor descriptors, solubilities, and binding affinities of selected odorants previously reported to bind wildtype pOBP.[2,31]

| Odorant | Odor Descriptors[a] | Solubility [mg/L][a] | Binding affinity [mu M][b] | Ref.[c] |
|---|---|---|---|---|
| Benzophenone | sweet, rosy, herbal | 137 | 3.6 ± 0.6 | [31] |
| Benzylbenzoate | balsamic, herbal | 15.39 | 3.9 ± 0.8 | [31] |
| Dihydromyrcenol | citrus, bergamot | 252.2 | 0.8 ± 0.1 | [31] |
| Undecanal | floral, waxy, citrus | 14.27 | 0.7 ± 0.2 | [31] |
| *(R)*-Carvone | spearmint, herbal | 1310 | 0.39/0.42[d] | [2] |
| *(S)*-Carvone | caraway, minty | 1300 | 0.39/0.42[d] | [2] |
| *(R)*-Menthol | minty-fresh | 456 | 1.2/1.8[d] | [2] |
| *(S)*-Menthol | minty-fresh | 456 | 1.2/1.8[d] | [2] |
| Thymol | thyme-like, medicinal | 900 | 2.5 ± 0.4 | [31] |

[a] Odor descriptions and solubilities (in water at 25 °C) are given as provided on http://www.thegoodscentscompany.com/ (04.08.2024). [b] IC50 values. For *(R)/(S)*-carvone and Menthol the $1/K_D$ values are provided.[2] [c] references for the binding affinities. [d] Values estimated from data in ref[2].

**Table 2.** Relative binding free energy values (in kJ/mol) of the carvone-pOBP complex obtained with the volume-based metadynamics method using a radius of the restraining sphere of 28 Å. The corresponding free energy landscape is shown in Figure 4. The values have been corrected for the change in translational entropy caused by the restraining potential (see Section III of the SI for details).

| [a] | $\Delta G_{MetaD}$ | $T\Delta S$[b] |
|---|---|---|
| *(R)*-Carvone | -18.16 | 75.44 |
| *(S)*-Carvone | -18.03 | 71.71 |

[a] Relative energies are given with respect to the unbound state set to 0.00 kJ/mol. [b] Transition state joining the bound and unbound state (see Figure 4).

### *Future Perspectives: Paving the Way towards Biosensors and Artificial Olfaction*

Finally, our study provides new atomistic insight on the transporter function of OBPs which constitute the very first step in the biological process of olfaction. This is a stepping stone that paves the way to fully understand the complex molecular action mechanisms underlying the olfactory sense. In the long run, this is crucial to exploit this challenging biological process for reverse engineering to to develop biosensors in artificial noses and to digitize the sense of smell. Due to their compact beta-barrel structure and the resulting exceptional stability towards thermal and chemical denaturation, OBPs seems to be the ideal candidates for application in artificial noses.[42,43] Although the binding of the odorants inside their binding pocket is non-specific,[31] it is possible to tune the physical properties of the binding pocket through point mutations to enhance the binding specificity towards selected molecules. While wild-type pOBP exhibits no binding preferences towards any of the carvone enantiomers, its F88W mutant was shown to successfully distinguish between the *(R)*- and *(S)*-enantiomer of carvone, and was used as a first bio-electronic capacitance-modulated transistor.[4] A recent study further showed that isoleucines within the pOBP binding pocket are crucial for the chiral discrimination of menthol and carvone.[2] Therefore, OBPS have moved more and more into the center of attention as potential odorant and biomarker detectors.[4,44] However, to exploit these protein targets, further understanding their structure and dynamics is essential to tune their physical properties, i.e., the binding affinity and specificity, towards detecting targeted biomarkers and relevant odorant families. Here, our work provides new insight into the mechanisms of the odorant uptake through OBPs. In future, it will required to step beyond targeting singled-out OBP-odorant complexes and adapt more global approaches that comprise larger protein data pools. With the release of alphafold structures[45] and sophisticated algorithms such as AlphaFill,[46] larger protein structural databases are becoming available and will allow employ machine learning approaches to rationalize the process of protein design and fine-tune OBPs to selectively target relevant molecular family. In near future, odorant uptake and absorption into the body is likely to gain importance and further open new possibilities and challenges for interdisciplinary research and multi-scale approaches, including, but not limited to the study of the function and mechanisms of the bio-transformation of odorants.[47] In any case, quantifying and digitizing the olfactory sense will certainly remain an ambitious goal that requires the joint effort from all disciplines ranging from fragrance chemistry and structure-odor predictions,[48–50] to molecular and structural biology,[51] neuroscience,[52] evolutionary biology,[53] and psychology.[54]

## Acknowledgements

The authors would like to acknowledge the SURFsara compute cluster hosted by SURF and the BAZIS research cluster hosted by VU for the computational time and the provided technical support. M.P. would like to thank the Erasmus+ trainee-ship program for funds (personal grant M.P.).

## Conflict of Interest

The authors declare no conflict of interest.

**Keywords:** Odorant Binding • High-Resolution Spectroscopy • Carvone • Enhanced Sampling • Biosensors

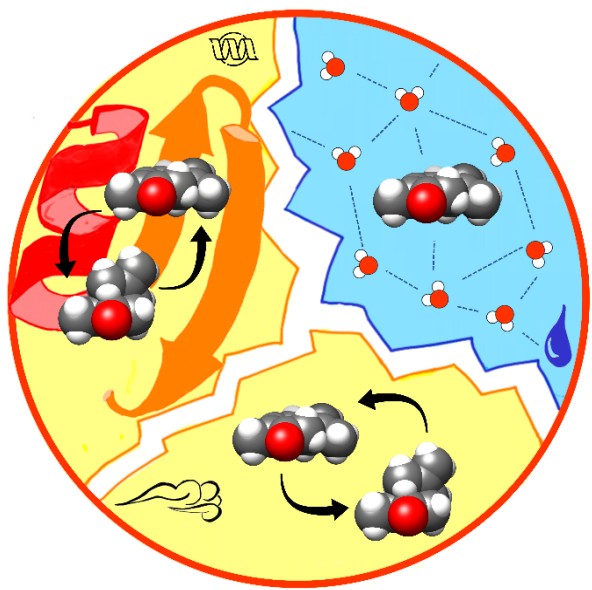

**Conformational freedom.** Porcine odorant binding protein mimics the gas phase environment of carvone by lowering the energy barrier of interconversion between different conformational states. A multi-scale computational approach reveals the driving force that allows OBPs to facilitate the transport of hydrophobic odorants through the hydrophilic mucus.

# Electronic Supporting information

## Mimicking the Gas-Phase to Transport Odorants through the Nasal Mucus: Functional Insights into Odorant Binding Proteins

Massimiliano Paesani,[a],[b] Arthur G. Goetzee,[a] Sanne Abeln,[a],[c] Halima Mouhib,*[a]

[a] Department of Computer Science, Bioinformatics, Vrije Universiteit Amsterdam, De Boelelaan 1105, 1081 HV Amsterdam, The Netherlands.
[b] Van't Hoff Institute for Molecular Sciences, Universiteit van Amsterdam, Science Park 904, 1090 GD Amsterdam, The Netherlands.
[c] Department of Computer Science, Universiteit Utrecht, Heidelberglaan 8, 3584 CS Utrecht, The Netherlands

*corresponding author: h.mouhib@vu.nl

**Content**

**Part 1 - Quantum Chemical Calculations**
**Part 2 - Molecular Dynamics**
**Part 3 - Umbrella Sampling**
**Part 4 - Volume-based Metadynamics**

**References**

# Part 1 - Quantum Chemical Calculations

The initial structures of *(R)*- and *(S)*-Carvone were generated manually and optimized at the B3LYP/6-311++g(d,p) level of theory using density functional theory and at the MP2/6-311++g(d,p) level of theory using the GAUSSIAN16 programme package.[FRI 2016] Harmonic frequency calculations were carried out to verify the nature of the stationary points and include the zero-point energy corrections. It should be noted that enantiomers are mirror images of one another and exhibit the same rotational constant and absolute (relative) energy differences. Therefore, they cannot be distinguished using the experimental set up used by Huet *et al*. [MOR 2013]. Nonetheless, For the sake of completeness, and to include both enantiomers during our molecular dynamics simulations, we treated both Carvone enantiomers, *(R)*- and *(S)*-Carvone, in our study.

Altogether, six stable energy minima were identified as local minima on the potential energy surface. These conformers can be globally classified into two subsets based on the equatorial or axial orientation of the isopropenyl group. Figure S1 and S2 depict the six conformers of *(R)*-Carvone and *(S)*-Carvone optimized at the MP2/6-311++(Gd,p) level of theory. The corresponding sum of electronic energies in Hartree (including the zero-point energy corrections), rotational constants, and relative energies in kJ/mol are given in Table S1 and Table S2. The numbering of the six conformers corresponds to the relative energies at the MP2/6-311++G(d,p) level of theory. As described in earlier works, conformers 1 and 3 are the most abundant conformers observed under supersonic jet conditions [URE 2009, MOR 2013]. Therefore, we used the optimized geometries of the lowest energy conformer (conformer 1) of *(R)*- and *(S)*-Carvone to subsequently set up and carry out the molecular dynamics simulations of both ligands *in vacuo*, in explicit water, and within the pOBP binding pocket as described in Section II *Molecular Dynamics* below.

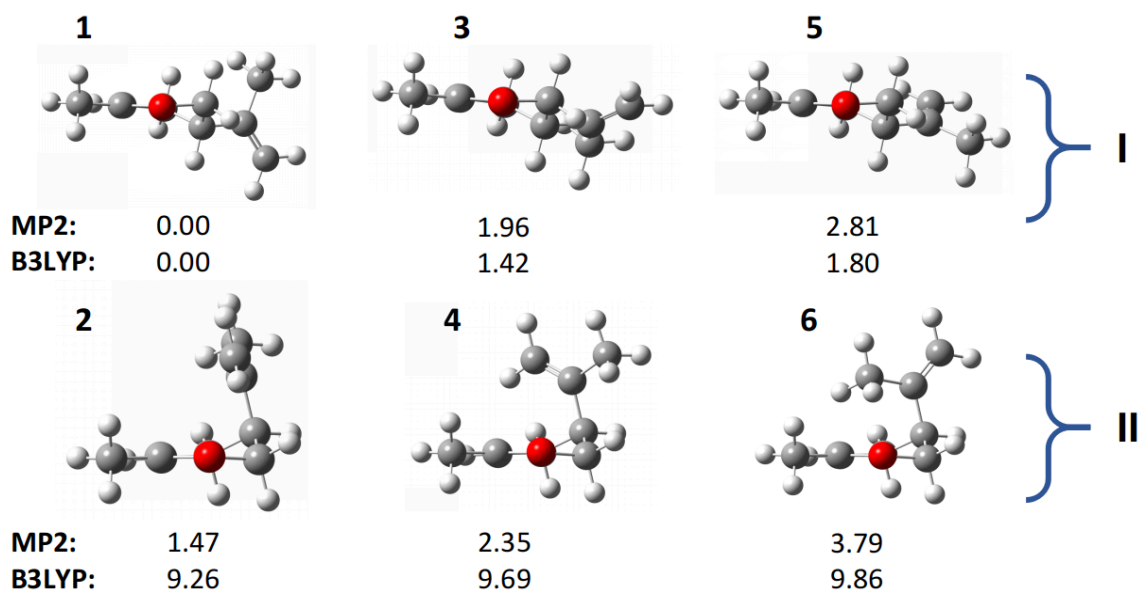

Figure S1. Optimized structures of *(R)*-Carvone obtained at the MP2/6-311++G(d,p) level of theory. The relative energies including the zero-point energy corrections and are given with respect to the global minimum (Conformer 1, cf. Table S1). Note that conformers 1, 3, and 5 have an equatorial orientation of the isopropenyl group (Set I), while conformers 2, 4, and 6 exhibit an axial orientation (set II). Conformers 1 and 3 are the most abundant conformations under experimental conditions (see ref.[MOR 2013]).

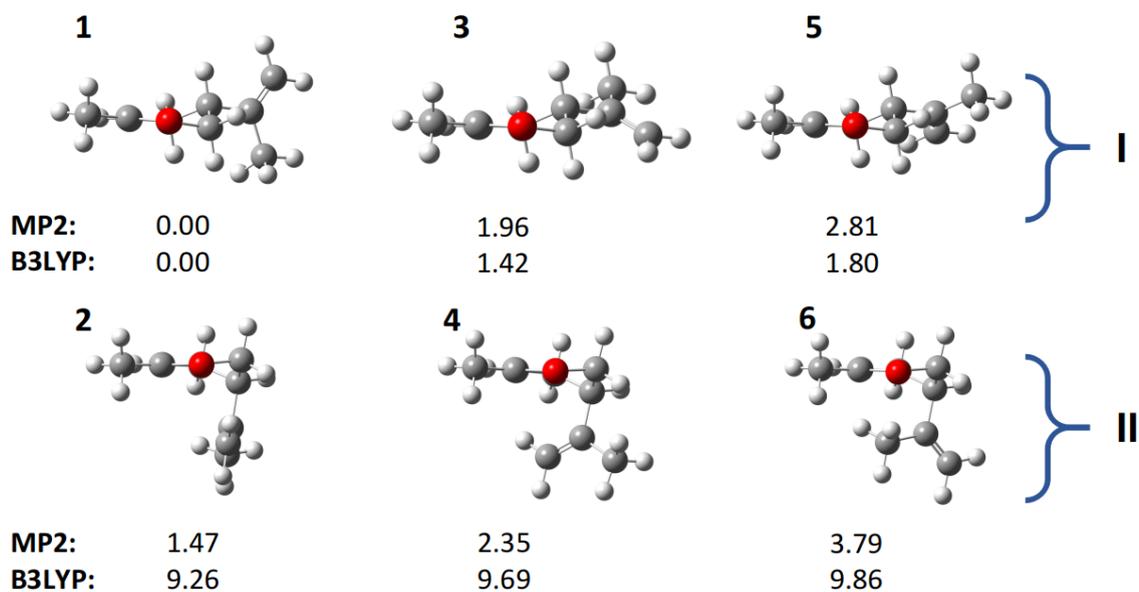

**Figure S2.** Optimized structures of *(S)*-Carvone obtained at the MP2/6-311++G(d,p) level of theory (cf. caption Fig. S1 for details).

**Table S1.** Quantum chemical Results obtained for the (R)-Carvone enantiomer (*cf.* FigureS1 for the corresponding structures at the MP2 level of theory).

| Conf.[a] | A[b] | B[b] | C[b] | µ[b] | E[b] | $E_{rel}$[b] | ZPE[b] | $ZPE_{rel}$[b] |
|---|---|---|---|---|---|---|---|---|
| B3LYP/6−311++G(d,p) | | | | | | | | |
| 1 | 2.237 | 0.651 | 0.575 | 3.61 | −464.825380 | 0.00 | −464.610376 | 0.00 |
| 2 | 1.739 | 0.840 | 0.734 | 3.96 | −464.822084 | 8.65 | −464.606849 | 9.26 |
| 3 | 2.258 | 0.666 | 0.550 | 3.82 | −464.824824 | 1.46 | −464.609836 | 1.42 |
| 4 | 1.669 | 0.860 | 0.744 | 3.47 | −464.821844 | 9.28 | −464.606685 | 9.69 |
| 5 | 2.214 | 0.676 | 0.552 | 3.25 | −464.824711 | 1.76 | −464.609690 | 1.80 |
| 6 | 1.823 | 0.808 | 0.741 | 3.48 | −464.821774 | 9.47 | −464.606619 | 9.86 |
| MP2/6−311++G(d,p) | | | | | | | | |
| 1 | 2.226 | 0.658 | 0.581 | 4.04 | −463.425216 | 0.00 | −463.208058 | 0.00 |
| 2 | 1.618 | 0.921 | 0.810 | 4.33 | −463.425164 | 0.14 | −463.207499 | 1.47 |
| 3 | 2.247 | 0.674 | 0.557 | 4.26 | −463.424473 | 1.95 | −463.207310 | 1.96 |
| 4 | 1.582 | 0.941 | 0.806 | 3.78 | −463.424635 | 1.53 | −463.207164 | 2.35 |
| 5 | 2.196 | 0.681 | 0.562 | 3.73 | −463.424130 | 2.85 | −463.206989 | 2.81 |
| 6 | 1.779 | 0.843 | 0.775 | 3.91 | −463.424026 | 3.13 | −463.206613 | 3.79 |

[a] Conformer labels given with respect to increasing relative energies obtained at the MP2/6-311++G(d,p) level of theory.
[b] A, B, C rotational constants in GHz; µ dipole moment in Debye; E, ZPE sum of electronic energies and sum of electronic energies inclusind zero-point energy correction in Hartree; $E_{rel}$, $ZPE_{rel}$, corresponding relative energies with respect to Conformer 1 in kJ/mol.

**Table S2.** Quantum chemical Results obtained for the (S)-Carvone enantiomer (*cf.* FigureS2 for the corresponding structures at the MP2 level of theory).

| Conf.[a] | A[b],[c] | B[b],[c] | C[b],[c] | μ[b] | E[b] | E$_{rel}$[b] | ZPE[b] | ZPE$_{rel}$[b] |
|---|---|---|---|---|---|---|---|---|
| colspan="9" | B3LYP/6-311++G(d,p) |
| 1 | 2.237 | 0.651 | 0.575 | 3.61 | −464.825380 | 0.00 | −464.610376 | 0.00 |
| 2 | 1.739 | 0.841 | 0.734 | 3.96 | −464.822084 | 8.65 | −464.606848 | 9.26 |
| 3 | 2.258 | 0.666 | 0.550 | 3.82 | −464.824824 | 1.46 | −464.609836 | 1.42 |
| 4 | 1.669 | 0.861 | 0.744 | 3.47 | −464.821844 | 9.28 | −464.606684 | 9.69 |
| 5 | 2.214 | 0.676 | 0.552 | 3.25 | −464.824711 | 1.76 | −464.609690 | 1.80 |
| 6 | 1.823 | 0.808 | 0.741 | 3.48 | −464.821774 | 9.47 | −464.606619 | 9.86 |
| colspan="9" | MP2/6-311++G(d,p) |
| 1 | 2.226 | 0.658 | 0.581 | 4.04 | −463.425220 | 0.00 | −463.208057 | 0.00 |
| 2 | 1.618 | 0.921 | 0.810 | 4.33 | −463.425160 | 0.16 | −463.207499 | 1.47 |
| 3 | 2.247 | 0.674 | 0.557 | 4.26 | −463.424470 | 1.97 | −463.207309 | 1.96 |
| 4 | 1.582 | 0.941 | 0.806 | 3.78 | −463.424630 | 1.55 | −463.207163 | 2.35 |
| 5 | 2.196 | 0.681 | 0.562 | 3.73 | −463.424130 | 2.86 | −463.206988 | 2.81 |
| 6 | 1.780 | 0.842 | 0.775 | 3.91 | −463.424030 | 3.12 | −463.206612 | 3.79 |

[a] Conformer labels given with respect to increasing relative energies obtained at the MP2/-311++G(d,p) level of theory.
[b] A, B, C rotational constants in GHz; μ dipole moment in Debye; E, ZPE sum of electronic energies and sum of electronic energies inclusind zero-point energy correction in Hartree; E$_{rel}$, ZPE$_{rel}$, corresponding relative energies with respect to Conformer 1 in kJ/mol.
[c] Note that the rotational constants do not differ from the corresponding (R)-Carvone enantiomers (cf. TableS1).

## Part 2 - Molecular Dynamics

### 2.1 Systems Preparation

The GROMACS program package version 2020.3 was used to perform the molecular dynamics (MD) simulations. Simulations were performed with the latest AMBER ff19SB-IDLN [TIA 2020]. at a temperature of 300K. For all molecular systems described below, 10 simulations of 50 ns each were carried out using an integration step of 2 fs. A different random seed was generated for each run to maximize the sampling of the respective configurational space. For the protein simulations the crystal structure of pOBP was downloaded from the Protein Data Bank (PDB-ID: 1A3Y)[SPI 1998]. Figure S3 shows the secondary structure of pOBP colored by the hydrophobicity of the amino acids residues.

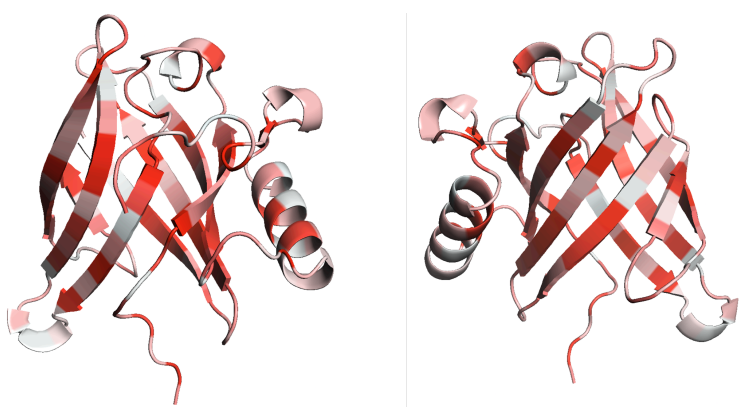

**Figure S3.** Crystal structure of pOBP (PDB-ID: 1A3Y [SPI 1998]) color coded by hydrophobicity, where red indicates stronger hydrophobicity. The applied scale for the hydrophobicity visualization using Pymol is based on ref.[EIS 1984].

**pOBP in explicit water.** The protein was solvated using the explicit water model TIP3P [PRI 2004] in a dodecahedral box using approximately 15,000 water molecules to increase the computational efficiency. The distance between the protein of the edges of the box was set to 1 nm. The system was neutralized using 14 Na$^+$ counter ions. After equilibration using the steepest descent algorithm [WAR 1988], the equilibration steps were performed using the canonical NVT (constant number of particles, volume and temperature) and the isothermal-isobaric NPT (constant number of particles, pressure and temperature) ensembles with position restraints to all heavy atoms using the LINCS algorithm during 100 ps each for both, the temperature, and the pressure equilibration steps. For the MD production runs, periodic boundary conditions were applied. The long-range electrostatic interactions were calculated using the Particle Mesh-Ewald (PME) method with a Fourier grid spacing of 0.12 nm. The cutoff for the short range electrostatic and van der Waals interactions was set to 1.0 nm. All bond lengths were constrained by the LINCS algorithm. The constant temperature and pressure conditions (T = 300 K, p = 1 bar) were ensured by coupling the system with a Nosé-Hoover thermostat [EVA 1985] and an Andersen-Parrinello-Rahman barostat [PAR 1981].

**pOBP-Carvone complex in explicit water.** To position the conformer inside the binding cavity of pOBP, *in silico* blind docking experiments were performed with Autodock4 and Autodock Tools [MOR 2009]. Using rigid side-chains and including a grid around the entire protein, the integrated option to combine an empirical free energy force field with a Lamarckian Genetic Algorithm [MOR 1998] was used for a total of 60 runs. The best scoring docking results based on lowest energy docked conformation for each cluster were used as a starting point to perform protein complex simulations. The production run of the complexes was subsequently performed analogous to the pOBP in explicit water (see description above). The parametrization of the ligand for the simulations is described below.

**LIGAND.** For the ligand simulations, the *(R)*- and the *(S)*-Carvone enantiomers were parameterized using the AMBER Force Field [PON 2003] with partial atomic charges as provided by the ATB03 web server [STR 2018]. The charges are given in Table S3 and Table S4 below. All protein-ligand complex simulations were carried out for both enantiomers using the starting geometry of the most abundant conformer (conformer 1) observed under experimental conditions by Huet *et al.*[MOR 2013] (see Section I *Quantum Chemical calculations* above for details). For the MD production runs, periodic boundary conditions were applied. The long-range electrostatic interactions were calculated using the Particle Mesh-Ewald (PME) method with a Fourier grid spacing of 0.12 nm. The cutoff for the short range electrostatic and van der Waals interactions was set to 1.0 nm . All bond lengths were constrained by the LINCS algorithm.

*In vacuo*. Gas-phase dynamical simulations were carried out at 300K. The structures were placed in an empty cubic box at a distance of 1nm from the edges. 10 simulations were performed for 50 ns for both carvone enantiomers. Energy minimization was performed with a steep descent algorithm for 50 ps.

*In explicit water.* Each enantiomer was solvated using the explicit water model TIP3P [PRICE 2004] in a cubic box using approximately 1150 water molecules. Periodic boundary conditions were applied. 10 simulations were performed for 50 ns for both conformers. Energy minimization was performed with a steep descent algorithm for 50 ps. Equilibration steps were performed using the canonical NVT (constant number of particles, volume and temperature) and the isothermal-isobaric NPT. The long-range electrostatic interactions were calculated using the Particle Mesh-Ewald (PME) method with a Fourier grid spacing of 0.12 nm. The cutoff for the short range electrostatic and van der Waals interactions was set to 1.0 nm. All bond lengths were constrained by the LINCS algorithm.

**Table S3**. Cartesian coordinates and partial charges of (*S*)-Carvone used for the MD production runs, as described under the Subsection above.

| Atom | x | y | z | charge |
| --- | --- | --- | --- | --- |
| H14 | 2.602 | -1.400 | -1.558 | 0.105 |
| C7 | 2.945 | -0.407 | -1.237 | -0.348 |
| H12 | 4.039 | -0.407 | -1.260 | 0.105 |
| H13 | 2.591 | 0.304 | -1.995 | 0.105 |
| C9 | 2.431 | -0.065 | 0.144 | 0.199 |
| C10 | 3.267 | 0.093 | 1.177 | -0.568 |
| H2 | 4.344 | -0.006 | 1.062 | 0.199 |
| H3 | 2.904 | 0.327 | 2.176 | 0.177 |
| C1 | 0.932 | 0.115 | 0.334 | 0.165 |
| H4 | 0.741 | 0.165 | 1.415 | 0.042 |
| C2 | 0.093 | -1.050 | -0.230 | -0.475 |
| H5 | 0.395 | -2.014 | 0.192 | 0.126 |

| Atom | x | y | z | charge |
|---|---|---|---|---|
| H6 | 0.227 | -1.116 | -1.321 | 0.149 |
| C8 | -1.396 | -0.861 | 0.018 | 0.649 |
| O1 | -2.137 | -1.825 | 0.204 | -0.559 |
| C5 | -1.930 | 0.522 | -0.029 | -0.052 |
| C6 | -3.424 | 0.684 | 0.086 | -0.395 |
| H9 | -3.708 | 1.740 | 0.042 | 0.122 |
| H10 | -3.792 | 0.262 | 1.028 | 0.122 |
| H11 | -3.942 | 0.149 | -0.720 | 0.122 |
| C4 | -1.080 | 1.558 | -0.187 | -0.087 |
| H1 | -1.491 | 2.566 | -0.258 | 0.141 |
| C3 | 0.416 | 1.440 | -0.268 | -0.279 |
| H7 | 0.722 | 1.531 | -1.322 | 0.128 |
| H8 | 0.879 | 2.291 | 0.248 | 0.107 |

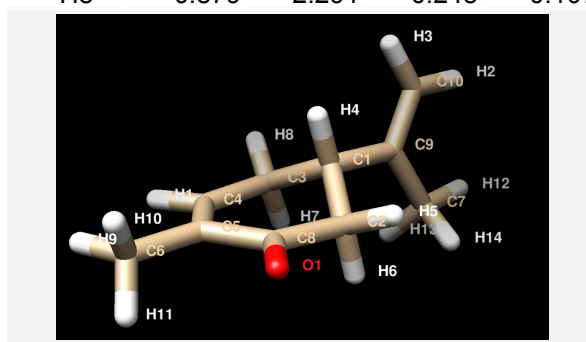

**Table S4**. Cartesian coordinates and partial charges of (*R*)-Carvone used for the MD production runs, as described under the Subsection above.

| Atom | x | y | z | charge |
|---|---|---|---|---|
| H14 | -4.299 | -0.438 | 1.108 | 0.195 |
| CAA | -3.225 | -0.304 | 1.219 | -0.556 |
| H13 | -2.830 | -0.372 | 2.231 | 0.175 |
| CAH | -2.432 | -0.071 | 0.166 | 0.184 |
| CAB | -2.997 | 0.019 | -1.234 | -0.310 |
| H10 | -2.812 | 1.004 | -1.684 | 0.095 |
| H11 | -4.078 | -0.149 | -1.232 | 0.095 |
| H12 | -2.542 | -0.720 | -1.906 | 0.095 |
| CAK | -0.933 | 0.099 | 0.356 | 0.147 |
| H9 | -0.730 | 0.034 | 1.434 | 0.040 |
| CAG | -0.111 | -1.013 | -0.329 | -0.398 |
| H1 | -0.414 | -2.008 | 0.012 | 0.106 |
| H2 | -0.268 | -0.983 | -1.418 | 0.131 |
| CAJ | 1.384 | -0.863 | -0.087 | 0.615 |
| OAD | 2.117 | -1.849 | -0.026 | -0.555 |
| CAI | 1.931 | 0.511 | 0.028 | -0.035 |
| CAC | 3.426 | 0.642 | 0.168 | -0.398 |
| H3 | 3.943 | 0.175 | -0.679 | 0.122 |
| H4 | 3.781 | 0.129 | 1.070 | 0.122 |
| H5 | 3.726 | 1.693 | 0.224 | 0.122 |
| CAE | 1.092 | 1.566 | -0.011 | -0.112 |
| H6 | 1.512 | 2.571 | 0.033 | 0.147 |
| CAF | -0.405 | 1.470 | -0.116 | -0.254 |
| H7 | -0.697 | 1.654 | -1.162 | 0.126 |
| H8 | -0.867 | 2.276 | 0.468 | 0.195 |

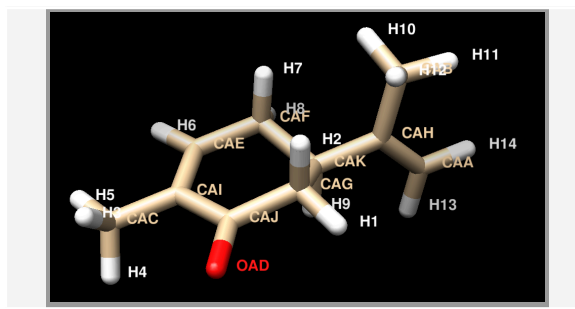

## 2.2 Data Analysis

The trajectories of the MD production runs were analyzed using available tools in GROMACS: 2020.3 and 2021.4.5 [ABR 2015] patched with plumed 2.7.2 [TRI 2014] to perform Volumed based metadynamics (see Section 4 further below). For visualization and making the figures, we used Chimera (https://www.cgl.ucsf.edu/chimera/), PyMol (https://www.pymol.org/), and VMD (https://www.ks.uiuc.edu/Research/vmd/).

**pOBP in explicit water: apo- and holo-form.**
To verify the impact of carvone on the size of the binding cavity, Figure S4 shows the RMSD distribution of pOBP in its apo-form and holo-form (including (R)- or (S)-carvove in the binding cavity) for 10 simulations of 50 ns of each system. The RMSD values were computed using the GROMACS g_rmsd tool.

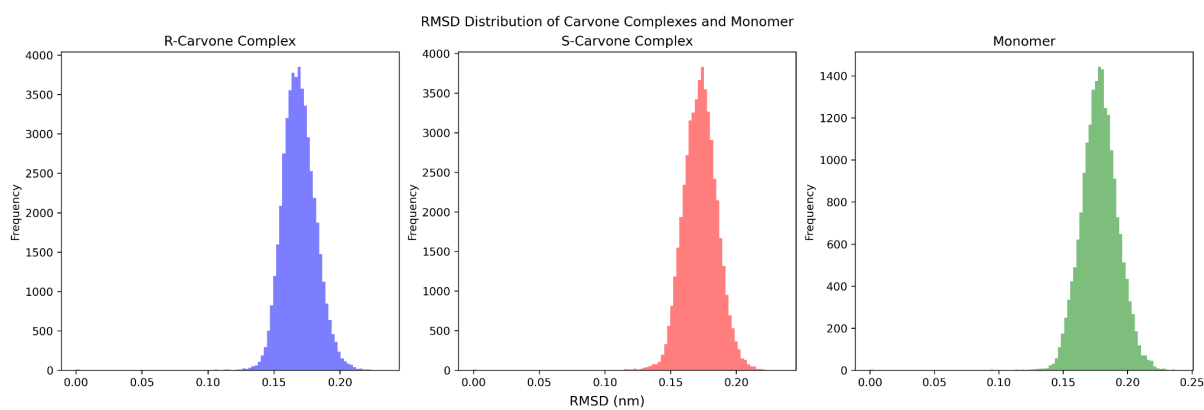

**Figure S4.** RMSD distribution of pOBP over all the possible states. Similar configurations are assumed by the protein, which maintains one main structure. Different color coded lines represent multiple simulations with different random seeds for a total of 10 simulations in each system (R- and (S)-Carvone). |(D) RMSD (C-alpha compared to starting structures) of the three states of pOBP. Values are all similar and overlap in multiple timesteps. Simulations were performed for 50 ns with timesteps of 2 fs. Note that this is in agreement with previous reports in the literature [GOL 2006].

While in the case of the Carvone conformations the radius of gyration provides an insightful order parameter to evaluate the sampled conformational space (see below), the RMSD computed for pOBP does not explicitly provide insight on the size and shape of the binding cavity and the resulting distributions may therefore be ambiguous [KUF 2012]. Therefore, in addition to computing the RMSD, we define unique cross-barrel distances of pOBP as order parameters to monitor changes in the dimension of the pOBP beta-barrel structure during the simulations. This also allows us to directly compare the impact of Carvone on the barrel diameter. Herefore, we explicitly define 3 distances within a horizontal plane going through the center of mass of the protein. A schematic representation is shown in Figure S5 below. The residue labels and the experimental distances identified in the crystal structure (PDB-ID: 1A3Y)[SPI 1998] are given in Table S5.

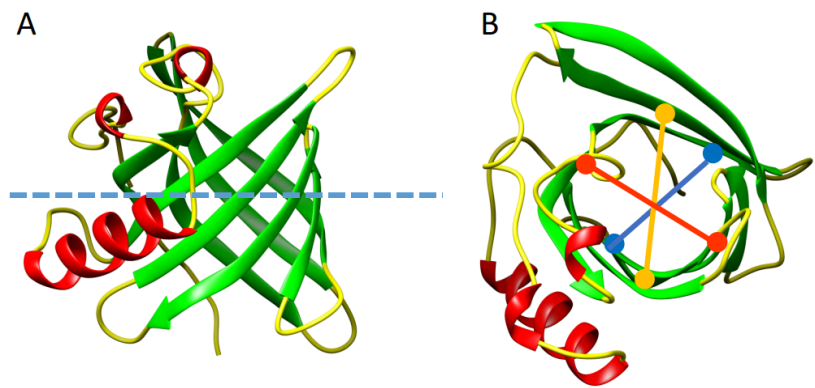

**Figure S5.** Graphical representation of the cross-distances defined as order parameters to monitor the dynamics of the beta-barrel structure during the simulations of pOBP in its *apo* and *holo* form. **A:** Side view of the protein on the beta-barrel structure. **B:** Top-view through the barrel structure of the protein (residue names and labels are given in Table S5).

**Table S5**. Residue pairs selected to define the cross-distances inside the binding cavity of pOBP as order parameter to monitor the dynamics of the beta barrel throughout the simulation. The standard deviation for the distances in the simulations is given in parentheses.

| Distance | aa pairs in pOBP | Crystal structure PDB-ID: 1A3Y [SPI 1998] | Complex pOBP (R)-Carvone | Complex pOBP (S)-Carvone | pOBP (apo-form) |
|---|---|---|---|---|---|
| **d1** | Lys15 - Asn104 | 2.38 | 2.38 (0.04) | 2.39 (0.04) | 2.39 (0.04) |
| **d2** | Ser57 - Ser91 | 2.29 | 2.38 (0.06) | 2.39 (0.06) | 2.40 (0.05) |
| **d3** | Asn54 - Ser101 | 1.71 | 1.74 (0.04) | 1.74 (0.04) | 1.75 (0.05) |

The cross-barrel distances depicted in Figure S6 below confirm that the protein remains unchanged by the insertion of carvone inside the binding cavity. This is in agreement with previous reports in the literature [GOL 2006] and supports the function of OBP as a universal carrier for hydrophobic molecules.

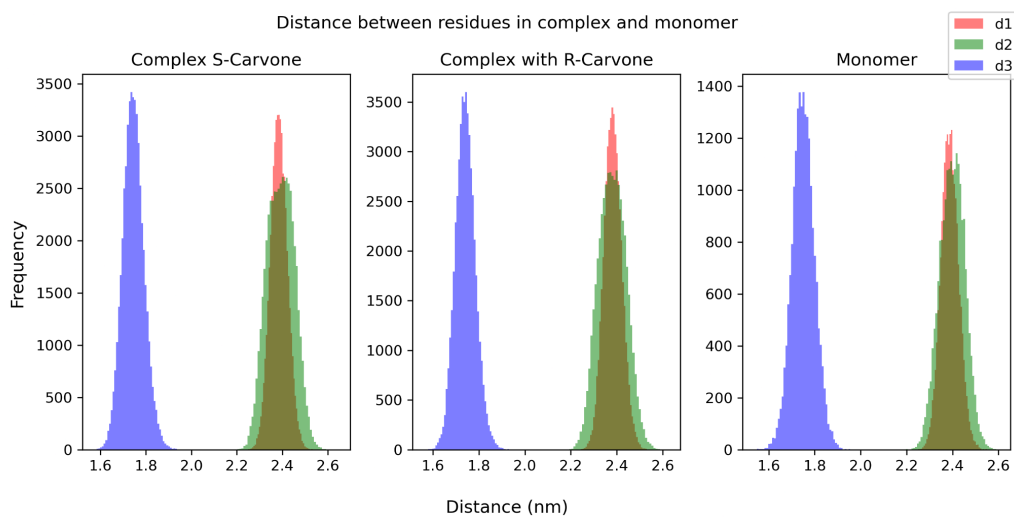

**Figure S6.** Distributions of the cross-distances inside the binding cavity of pOBP: in complex with (S)-Carvone and (R)-Carvone, and as a monomer in its apo-form. See Table S5 for the corresponding residue pairs and Figure S5 for a graphical representation.

**Carvone in its different environment: *in vacuo*, in explicit water, and the protein environment.**

To verify the conformational space explored by *(R)*- and *(S)*-carvone in the different simulations, the radius of Gyration of the odorant was first monitored in the three different states (see Figure S7). In addition, we defined a C-C intramolecular distance as an additional order parameter to quantify the conformational flexibility of the odorant (see Figure S8).

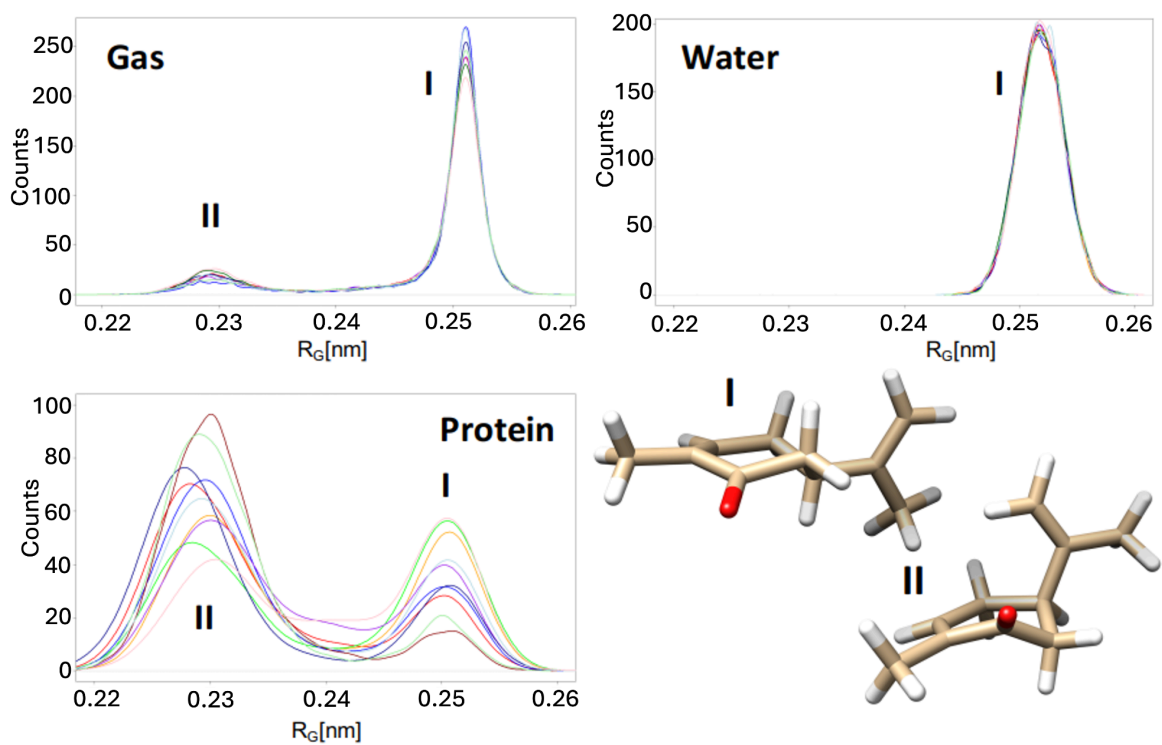

**Figure S7.** Frequency distribution over the simulated Radius of gyration of (R)-carvone in the gas, explicit water, and protein-ligand complex simulations (each graph contains 10 production runs of 50 ns with a different random seed). State I and II shows a representative snapshot of the (R)-carvone conformation indicated in the plots. Note that in explicit water, the simulations are not able to sample the different states of carvone. The sampling is only possible in the gas and inside the protein binding cavity which suggests a higher energy barrier to switch between the different states of the odorant in explicit water (see ms. from further details and Section 3 on umbrella sampling below to estimate the barrier height).

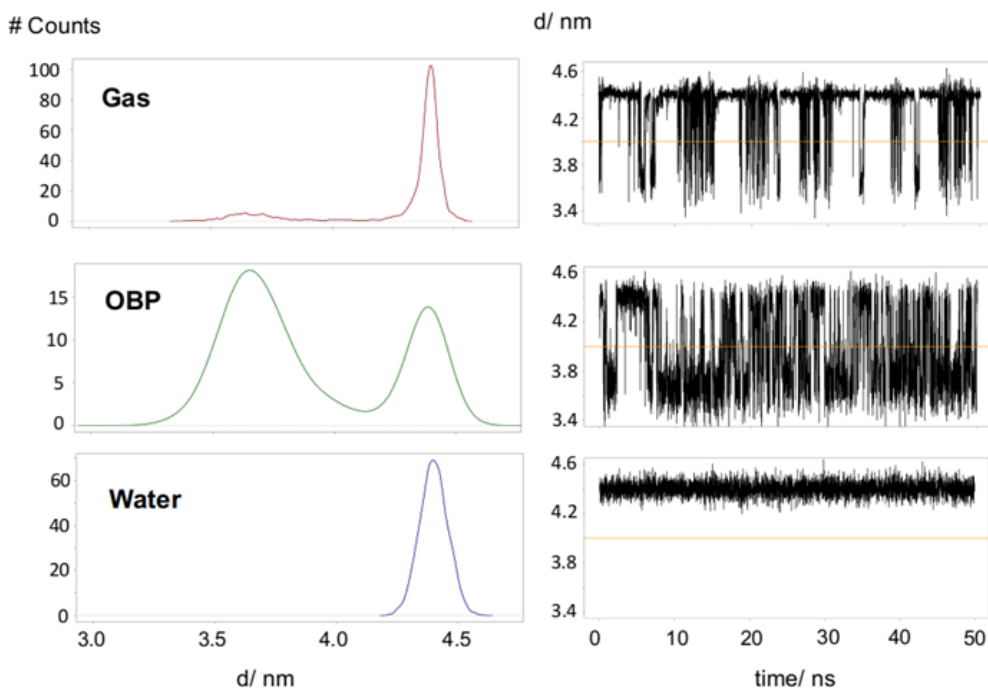

**Figure S8.** Frequency distribution over the C-C distance of (R)-carvone (see ms. for details) in the gas, explicit water, and protein-ligand complex simulations (simulation runs of 50 ns each). Note that in explicit water unlike in the other states, the simulations are not able to sample the different states of carvone within the simulation time. Therefore additional umbrella sampling simulations are required to estimate the energy barrier of transition between the two different states of carvone (see further below).

As can be seen in Figure S8 the two different conformational states of carvone cannot be sampled using a regular MD production run in explicit water. Here, umbrella sampling simulations are required to sample both states and estimate the free energy barrier for interconversion between the two states. Figure S9 further shows that sampling is also not achieved for a longer simulation time.

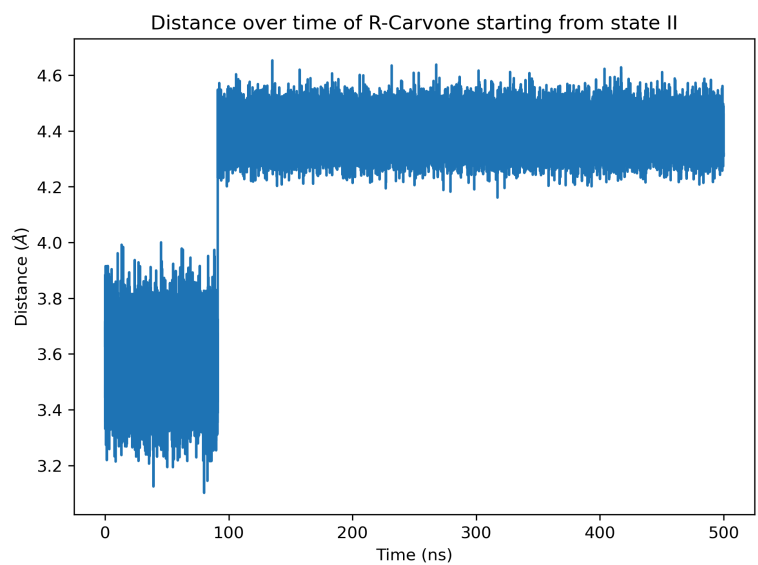

**Figure S9.** C-C distance of (R)-carvone (see ms. for definition of C-C distance) in the explicit water simulation using the conformational state II as starting point of the simulation (see FIgure S8 for simulations starting from state I). Note that this simulation run was performed for 500 ns and depicts that the two different conformational states cannot be sampled in explicit water using a regular MD production run.

## Part 3 - Umbrella Sampling

To estimate the barrier height for the transition of *(R)*-Carvone and *(S)*-Carvone in explicit water, umbrella sampling (US) simulations were performed using GROMACS 2021.5 [ABR 2015] patched with PLUMED 2.8.0 package [TRI 2014].

**3.1 Umbrella sampling**

The umbrella bias for simulations is defined by a harmonic potential $\frac{k}{2}(x-a)^2$, where x represents the US window in the order parameter space, $a$ represents the instantaneous value of the order parameter and $k$ represents the spring constant which defines the strength of the US bias in kJ/mol. In our system, the value of $k$ in was determined to be $1.0 \times 10^6$ kJ/mol by running a few short simulations and monitoring the value of the order parameter.

The starting coordinates for each US window were generated using a steered molecular dynamics simulation by using the structure of conformer 6 of (R)-carvone and pulling at a constant rate of 0.1 nm per ps on the carbon atoms C1 and C8 with a spring constant $k$ of 1000 kJ/mol, to elongate the C-C distance from 3.7 Å to 4.4 Å and cover the order to sample the transition between the two conformations in the US simulations. PDB-structures were extracted from the simulations and the US windows were defined around 0.32 nm to 0.46 nm with 0.01 nm increments. However, the transition state (between 0.39 nm to 0.42 nm) was, in addition to the regular windows, further upsampled with 0.005 nm increments and a larger value for $k$ of $1.0 \times 10^7$ kJ/mol. All umbrella windows and their value for $k$ can be found in Table S6. The same procedure was carried out for (S)-carvone.

Each US window was subsequently run for 20 ns (10000000 steps, dt=0.002 ps). Order parameter values were tracked and stored at every simulation step. The corresponding US windows and their respective spring constants are provided in Table S6 below.

**Table S6.** US windows and their respective spring constants.

| Distance (nm) | Spring constant $k$ (kJ/mol) |
|---|---|
| 0.32 | 100000 |
| 0.33 | 100000 |
| 0.34 | 100000 |
| 0.35 | 100000 |
| 0.36 | 100000 |
| 0.37 | 100000 |
| 0.38 | 100000 |
| 0.39 | 100000 |
| 0.395 | 1000000 |
| 0.40* | 100000, 1000000 |
| 0.405 | 1000000 |
| 0.41* | 100000, 1000000 |
| 0.415 | 1000000 |
| 0.42 | 100000 |
| 0.43 | 100000 |
| 0.44 | 100000 |
| 0.45 | 100000 |
| 0.46 | 100000 |

\* Distances with an asterisk (*) were sampled twice using a larger spring constant.

### 3.2 Data processing

Next, the order parameter samping was processed using the weighted histogram analysis method (WHAM) tool (version 2.0.10) developed by Grossfield *et al.* [GRO 2014] using 150 bins and a maximum tolerance of $1.0 \times 10^{-7}$.

### 3.3 Analysis

Figure S10 shows the histograms of the order parameter sampling for each US window. As can be seen from the figure, there is sufficient overlap for the WHAM method to be employed.

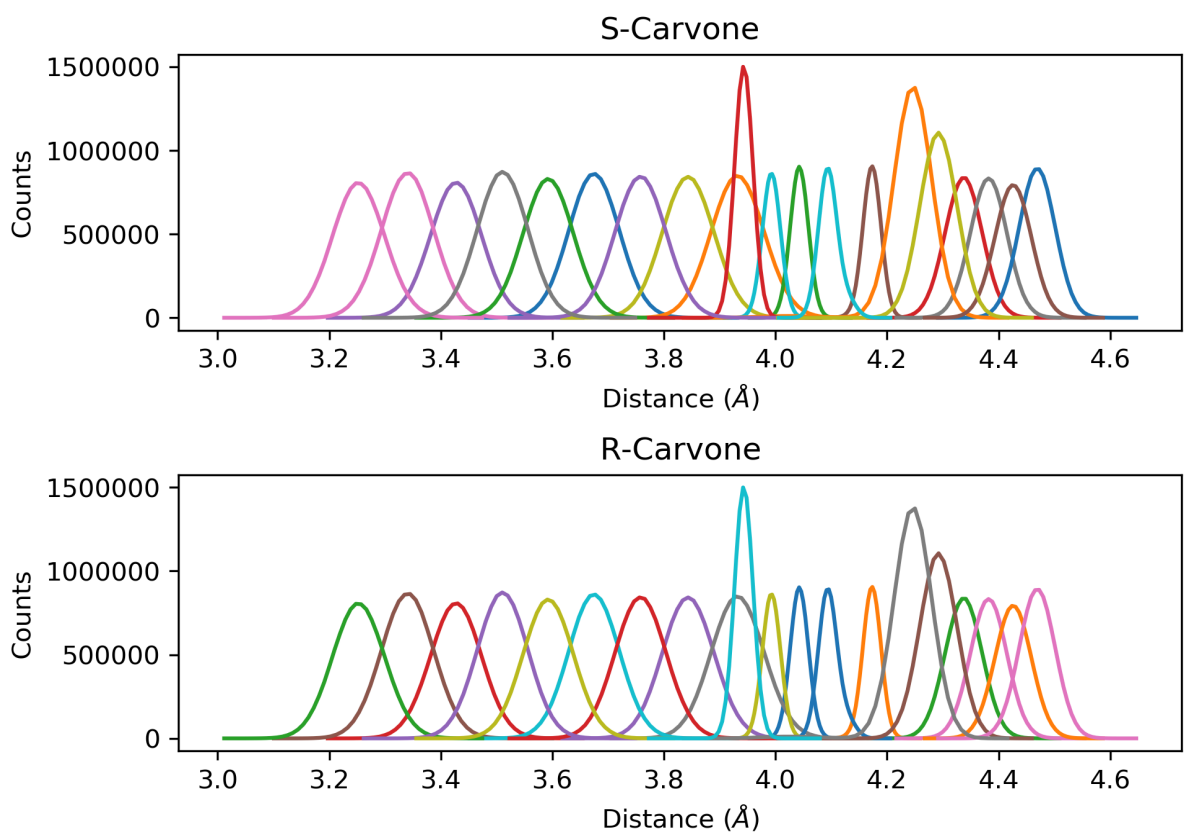

**Figure S10.** Order parameter sampling histograms of the US simulations for (S)-Carvone (top) and (R)-Carvone (bottom).

The free energy surface as a function of distance of *(S)*-Carvone in different environments is provided in Figure S11 below (with the free energy surfaces of (R)-carvone depicted as dotted lines). The free energy curve in explicit water is obtained from the US simulations described above. This is in agreement with FigureS9, which showed that due to the high barrier of interconversion between the two conformational states of carvone, molecular dynamics are not sufficient to accurately sample the different states and enhanced sampling techniques are required.

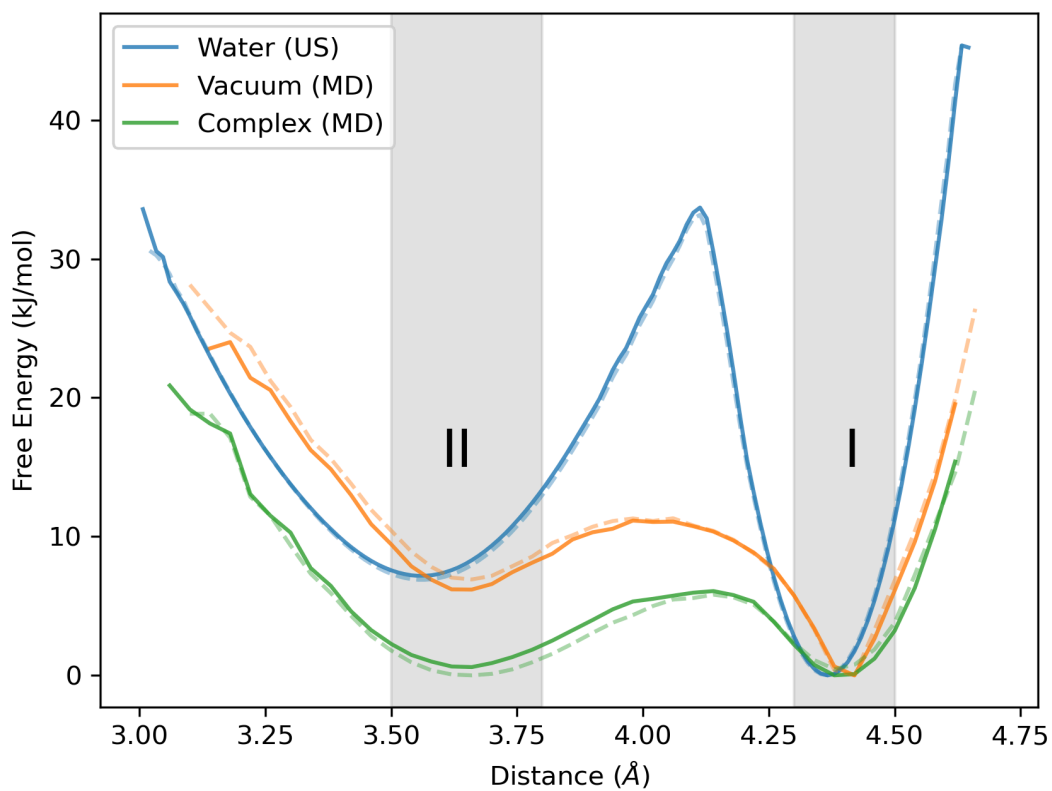

**Figure S11.** Free energy surface as a function of distance of *(S)*-Carvone in different environments showcasing the different states: water (hydrophilic mucus), vacuum(air), and in complex (protein environment) using umbrella sampling, *in vacuo* molecular dynamics, and explicit water molecular dynamics, respectively.

## Part 4 - Volume-based Metadynamics

### 4.1 Background

Volume Based Metadynamics were performed as described in the original work of Capelli *et al.* [CAP 2019]. Hereby, a sphere of finite radius $\rho_s$ is defined that needs to be larger than the radius of gyration of our protein in order to enclose its entire β-barrel structure. Within the defined volume, the spherical coordinates (ρ, θ, φ) are used as collective variables (CVs) and a repulsive harmonic potential is added at the border of the sphere to limit the volume in the solvated state that requires sampling. This harmonic potential is described as

$$U_s(\rho(t)) = \frac{1}{2}(\rho(t) - \rho_s)^2 \quad (if \ \rho(t) > \rho_s, \ otherwise = 0) \tag{1}$$

where the harmonic confinement constant k = 10 kJ/mol/Å² needs to be large enough to prevent the ligand from escaping the confining volume. Figure S12 below shows a schematic representation of the CVs and our protein inside the sphere. It is important that the radius of the sphere be larger than the radius of gyration of our protein to comprise the complete beta-barrel structure. The correction for the resulting change in the translational entropy of the solvated state based on applying the repulsive potential is given in the supporting information and has been described by the authors in the original work.

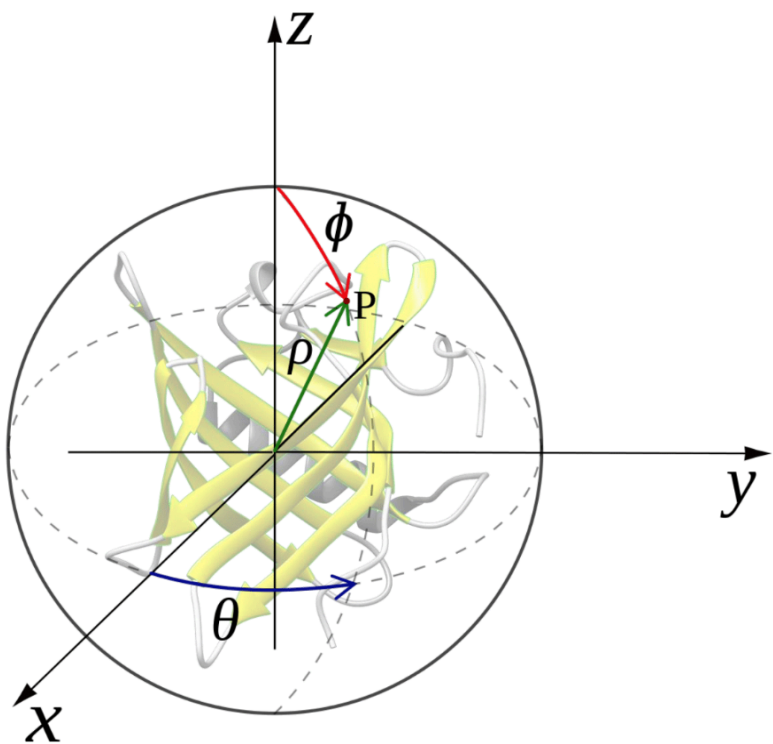

**Figure S12.** Schematic representation of the collective variables (CVs) used for the volume based metadynamics implementation. The three CVs that describe the position of the ligand in the simulations are the spherical coordinates ρ, Θ, and φ (Figure adapted from the Electronic Supporting Information provided by Capelli *et al.*).

Due to the restraining potential, a change in the translational entropy of the solvated state occurs and, for this reason, a correction to the binding free energy needs to be applied in this form:

$$\Delta G_b^0 = - RT\log(C^0 K_b) \qquad (2)$$

$$\Delta G_b^0 = \Delta G_{MetaD} - RT\log\left(\frac{V^0}{\frac{4}{3}\pi\rho_s^3 - V_{prot}}\right) \qquad (3)$$

with $R$ being the gas constant, $T$ the system temperature, $K_b$ the binding constant, $C_0$ = 1660 Å³ the standard concentration, $V^0$ its reciprocal, $\Delta G_{MetaD}$ the binding free energy obtained by well-tempered metadynamics, and $V_{prot}$ the volume of the protein inside the restraining potential. The derivation has been described in Electronic Supporting Information of the original work by Capelli et al. [CAP 2019]. Their approach fully takes the rotational entropy contribution into account as the presence of the spherical restraint does not act on the rotational degrees of freedom of the ligand, and the metadynamics approach does not affect the rotational entropy.

**4.2 Production Runs**
Simulations were performed using the GROMACS version 2021.5 patched with PLUMED 2.7.0.5, for both the (R)- and the (S)-Carvone enantiomer in complex with pOBP. The starting structures of the protein-ligand complexes (for both enantiomers) were set up as described under the Supporting Information Section '*Systems Preparation*' above. Simulations were run for a total of 200 ns for different radii of 20, 24, and 28 Å using a timestep of 2 fs. To guarantee the correct sampling of the bound and the unbound state, we limit the restraining volume of the sphere, i.e., of the 3 CVs to include the binding pose of ligand inside the binding cavity and enough volume to observe the ligand outside the protein, i.e., completely solvated. Therefore we sampled the free energy landscape of pOBP bound to (R)- and (S)-Carvone to compare and quantify the system in its bound and unbound state. We simulated 200 ns long production runs of enhanced sampling for the (R)- and (S)-Carvone pOBP complexes using different radii from the center of mass of the protein, at 20 Å, 24 Å, and 28 Å. The resulting free energy surface is then projected on two more intuitive CVs to describe the structure of the complex: the distance of Carvone from the center of mass of the protein, and the coordination number (CN, see equation 4), which captures the amount of non-covalent contacts between Carvone and the residues of the pOBP binding site.

**4.3 Data Analysis**
To evaluate the volume-based metadynamics, the calculated free energy is projected as a function of the sphere dimensions on two collective variables (CVs). This is described in the original work by Capelli *et. al.*, and we apply the same reweighting procedure.[TIW 2015] The applied CVs are the radius $\rho_s$ and the coordination number ($C_n$), which measures whether there is a contact between the ligand and the host protein using the following equation (see also ms.):

$$C_N = \sum_{i \in A} \sum_{j \in B} \frac{1 - (\frac{r_{ij}}{r_0})^n}{1 - (\frac{r_{ij}}{r_0})^m} \quad (4)$$

where $A$ and $B$ are the sets of non-hydrogen atoms of ligand and protein, $r_0$ is the threshold to define a contact (4.5 Å) and $r_{ij}$ is the distance between atoms $i$ and $j$, m = 6 and n = 12 as exponents of the switching function. An estimation of the absolute free energy was obtained by subtracting the energy minima (bound state) by the the energy of one of the assumed unbound states. The resulting free energy surfaces and plots are given in Figures S13 below (see also Figure 5 in the ms. for further details).

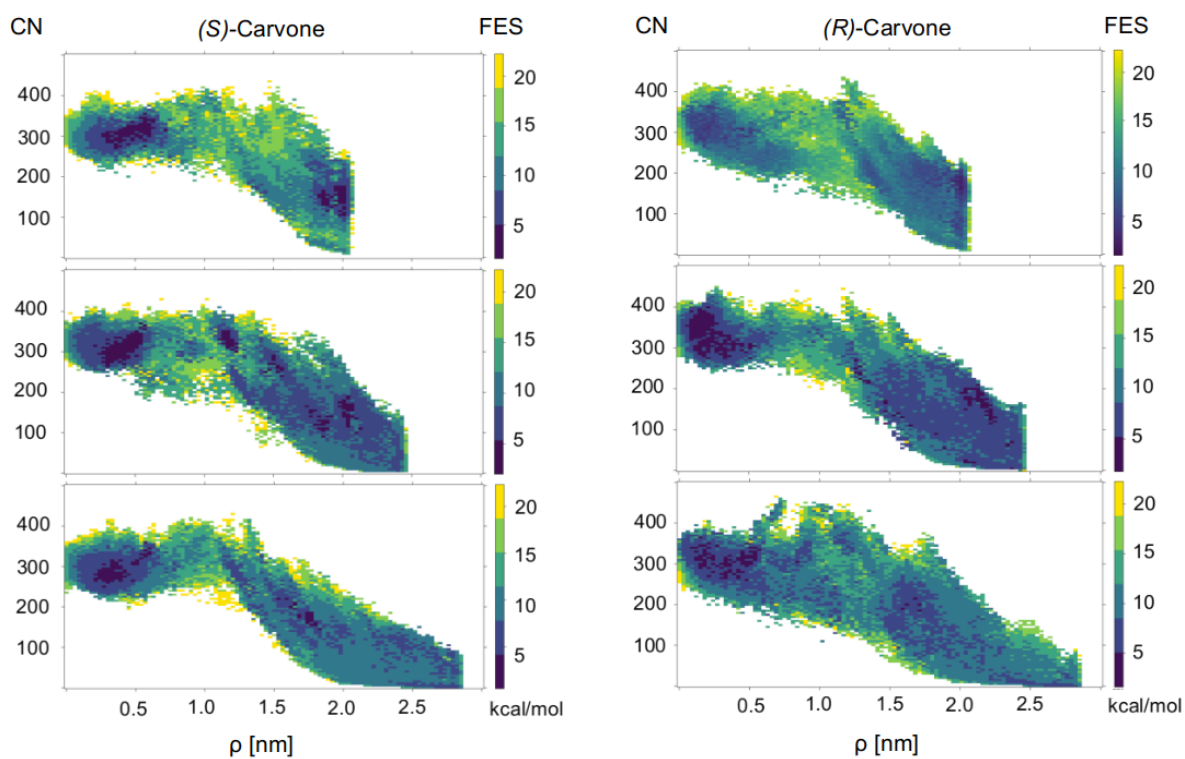

**Figure S13.** Free energy surfaces as a function of the restraint radius $\rho_s$ 20, 24 and 28 nm. **A:** (R)-Carvone-pOBP complex, **B:** (S)-Carvone-pOBP complex. CN: coordination number, FES: free energy surface. See ms. for further information.